\documentclass[aps,prd,twocolumn,superscriptaddress]{revtex4}
\usepackage{epsfig,epsf}
\usepackage{amsmath}
\usepackage{amsthm}
\usepackage{amsfonts}
\usepackage{amssymb}
\usepackage{dsfont}
\usepackage{multirow}
\usepackage{appendix}
\usepackage{slashed}
\usepackage[active]{srcltx}
\usepackage{psfrag}

\begin{document}

\title{Is the resonance $X_0(2900)$ a ground-state or radially excited
scalar tetraquark $[ud][\overline{c}\overline{s}]$? }
\date{\today}
\author{S.~S.~Agaev}
\affiliation{Institute for Physical Problems, Baku State University, Az--1148 Baku,
Azerbaijan}
\author{K.~Azizi}
\affiliation{Department of Physics, University of Tehran, North Karegar Avenue, Tehran
14395-547, Iran}
\affiliation{Department of Physics, Do\v{g}u\c{s} University, Acibadem-Kadik\"{o}y, 34722
Istanbul, Turkey}
\affiliation{School of Particles and Accelerators, Institute for Research in Fundamental
Sciences (IPM) P.O. Box 19395-5531, Tehran, Iran}
\author{H.~Sundu}
\affiliation{Department of Physics, Kocaeli University, 41380 Izmit, Turkey}

\begin{abstract}
We investigate properties of the ground-state and first radially excited
four-quark mesons $X_0$ and $X_0^{\prime}$ with a diquark-antidiquark
structure $[ud][\overline{c}\overline{s}]$ and spin-parities $J^{\mathrm{P}%
}=0^{+}$. Our aim is to reveal whether or not one of these states can be
identified with the resonance $X_0(2900)$, recently discovered by the LHCb
collaboration. We model $X_0$ and $X_0^{\prime}$ as tetraquarks composed of
either axial-vector or scalar diquark and antidiquark pairs. Their
spectroscopic parameters are computed by employing the QCD two-point sum
rule method and including into analysis vacuum condensates up to dimension $%
15$. For an axial-axial structure of $X_0^{(\prime)}$, we find partial
widths of the decays $X_0^{(\prime)} \to D^{-}K^{+}$ and $X_0^{(\prime)} \to
D^{0}K^{0}$, and estimate full widths of the states $X_0^{(\prime)}$. To
this end, we calculate the strong couplings at the vertices $%
X_0^{(\prime)}DK $ in the framework of the light-cone sum rule method. We
use also technical approaches of the soft-meson approximation necessary to
analyze tetraquark-meson-meson vertices. Obtained results $m=(2545 \pm 160)~%
\mathrm{MeV}$ and $m^{\prime}=(3320 \pm 120)~\mathrm{MeV}$ [$m_{\mathrm{S}%
}=(2663 \pm 110)~\mathrm{MeV}$ and $m_{\mathrm{S}}^{\prime}=(3325 \pm 85)~%
\mathrm{MeV}$ for a scalar-scalar current] for the masses of the particles $%
X_0$ and $X_0^{\prime}$, as well as estimates for their full widths $%
\Gamma_{0}=(140 \pm 29)~\mathrm{MeV}$ and $\Gamma_{0}^{\prime}=(110 \pm 25)~%
\mathrm{MeV}$ allow us to interpret none of them as the resonance $X_0(2900)$%
. At the same time, these predictions provide important information about
ground-state and radially excited diquark-antidiquark structures $X_0$ and $%
X_0^{\prime}$, which should be objects of future experimental and
theoretical studies.
\end{abstract}

\maketitle


\section{Introduction}

\label{sec:Int} 

One of important achievements of last years in physics of multiquark hadrons
is observation of structures $X_{0}(2900)$ and $X_{1}(2900)$ by the LHCb
collaboration. These resonance-like peaks were discovered in the invariant
mass distribution $D^{-}K^{+}$ of the decay channel $B^{+}\rightarrow
D^{+}D^{-}K^{+}$ \cite{LHCb:2020A,LHCb:2020}. The LHCb measured masses and
widths of these structures and fixed also their spin-parities. It turned
out, that $X_{0}(2900)$ and $X_{1}(2900)$ are the scalar and vector
resonances with quantum numbers $J^{\mathrm{P}}=0^{+}$ and $J^{\mathrm{P}%
}=1^{-}$, respectively.

Appearance of the mesons $D^{-}$ and $K^{+}$ at the final state of their
decays implies that $X_{0}(2900)$ and $X_{1}(2900)$ are composed of quarks $%
\overline{c}\overline{s}ud$, and may be considered as particles containing
four quarks of different flavors. In other words, $X_{0}(2900)$ and $%
X_{1}(2900)$ are presumably new evidences for exotic mesons with full
open-flavor structures. This is important fact, because existence of the
resonance $X(5568)$, presumably built of $sd\overline{b}\overline{u}$ quarks
and considered as a first candidate to fully open-flavor four-quark state
\cite{D0:2016mwd}, was not confirmed by other collaborations. Of course,
this analysis is correct in the context of the four-quark model of $%
X_{0}(2900)$ and $X_{1}(2900)$, because there are theoretical analyses which
claim to explain the LHCb data by hadronic rescattering effects. The LHCb
collaboration also did not exclude such interpretation of the observed
structures.

New experimental information triggered intensive theoretical activities
aimed to reveal internal organization of these resonances, calculate their
parameters and study processes in which $X_{0}(2900)$ and $X_{1}(2900)$ can
be produced \cite%
{Karliner:2020vsi,Wang:2020xyc,He:2020jna,Chen:2020aos,Liu:2020nil,Molina:2020hde,Hu:2020mxp, He:2020btl,Liu:2020orv,Lu:2020qmp,Zhang:2020oze,Huang:2020ptc,Xue:2020vtq, Yang:2021izl,Wu:2020job,Abreu:2020ony,Wang:2020prk,Xiao:2020ltm,Dong:2020rgs,Burns:2020xne, Bondar:2020eoa,Chen:2020eyu,Albuquerque:2020ugi}%
. In overwhelming majority of investigations, the resonances $X_{0}(2900)$
and $X_{1}(2900)$ were modeled as diquark-antidiquark states or hadronic
molecules. In fact, as a scalar tetraquark $[sc][\overline{u}\overline{d}]$
the resonance $X_{0}(2900)$ was explored in Refs.\ \cite%
{Karliner:2020vsi,Wang:2020xyc} using a phenomenological model and the sum
rule method, respectively. Predictions for the mass $(2863\pm 12)~\mathrm{MeV%
}$ and $(2910\pm 120)~\mathrm{MeV}$ obtained in these papers allowed the
authors to interpret $X_{0}(2900)$ as the ground-state scalar tetraquark $%
[sc][\overline{u}\overline{d}]$. An interesting assumption about nature of $%
X_{0}(2900)$ was made in Ref.\ \cite{He:2020jna}, where it was studied as a
radially excited state $[ud][\overline{c}\overline{s}]$. In the articles
\cite{Chen:2020aos,Liu:2020nil,Molina:2020hde,Hu:2020mxp} the resonance $%
X_{0}(2900)$ was examined as $S$-wave molecule $D^{\ast -}K^{\ast +}$. The
tetraquark and molecule models were used for the resonance $X_{1}(2900)$, as
well \cite{He:2020jna,Chen:2020aos,He:2020btl}. But two resonance-like peaks
in the $D^{-}K^{+}$ mass distribution may have alternative nature and emerge
due to triangle singularities in the rescattering diagrams $\chi
_{c1}D^{\ast -}K^{\ast +}$ and $D_{sJ}\overline{D}_{1}^{0}K^{0}$ \cite%
{Liu:2020orv}.

In Ref.\ \cite{Agaev:2020nrc}, we investigated $X_{0}(2900)$ as a molecule $%
\overline{D}^{\ast 0}K^{\ast 0}$ and evaluated its spectroscopic parameters
and width. Comparing our results for the mass $(2868\pm 198)~\mathrm{MeV}$
and width $(49.6\pm 9.3)~\mathrm{MeV}$ of $\overline{D}^{\ast 0}K^{\ast 0}$
with corresponding LHCb data $m=(2866\pm 7\pm 2)~\mathrm{MeV}$ and $\Gamma
=(57\pm 12\pm 4)~\mathrm{MeV}$, we decided a molecule model is acceptable
for the resonance $X_{0}(2900)$.

The vector resonance $X_{1}(2900)$ was considered in the context of the
diquark-antidiquark model in our article \cite{Agaev:2021knl}. We studied it
as a vector tetraquark built of a diquark $u^{T}C\gamma _{5}d$ and an
antidiquark $\overline{c}\gamma _{\mu }\gamma _{5}C\overline{s}^{T}$, and
computed relevant parameters. Though predictions for the mass $(2890\pm 122)~%
\mathrm{MeV}$ and width $(93\pm 13)~\mathrm{MeV}$ of this tetraquark are
smaller than the relevant LHCb data, we interpreted it as the resonance $%
X_{1}(2900)$ by keeping in mind that theoretical and experimental
investigations suffer from certain errors.

During last few years diquark-antidiquark states containing four quarks
(antiquarks) $c$, $s$, $u$ and $d$ in different configurations were objects
of investigations. Thus, a scalar tetraquark $X_{c}=[su][\overline{c}%
\overline{d}]$ was considered in our article \cite{Agaev:2016lkl}, where it
was modeled as an exotic meson made of scalar-scalar and axial-axial
diquarks with $C\gamma _{5}\otimes \gamma _{5}C$ and $C\gamma _{\mu }\otimes
\gamma ^{\mu }C$ type interpolating currents, respectively. The mass of $%
X_{c}$ found using these two structures is $(2634\pm 62)~\mathrm{MeV}$ and $%
(2590\pm 60)~\mathrm{MeV}$, respectively. The result $(2.55\pm 0.09)~\mathrm{%
GeV}$ for the mass of $X_{c}$ was obtained also in Ref.\ \cite{Chen:2016mqt}.

Though $X_{c}$ and $X_{0}=[ud][\overline{c}\overline{s}]$ have similar
content, there are two differences between them: $X_{c}$ is built of a
relatively heavy diquark $[su]$ and heavy antidiquark $[\overline{c}%
\overline{d}]$, whereas $X_{0}$ has a light diquark $[ud]$-heavy antidiquark
$[\overline{c}\overline{s}]$ structure. The second difference is decay
channels of these particles. While dominant decay mode of $X_{c}$ is $%
X_{c}\rightarrow D_{s}^{-}\pi ^{+}$, in the case of $X_{0}$ we have $%
X_{0}\rightarrow D^{-}K^{+}$. Nevertheless, as we shall see below, masses
and widths of $X_{0}$ and $X_{c}$ are close to each other mainly due to
their quark contents.

In the current work, we explore the scalar tetraquark $X_{0}=[ud][\overline{c%
}\overline{s}]$ in a detailed form. Thus, we compute masses of the
ground-state $1S$ and radially excited $2S$ tetraquarks $X_{0}$ and $%
X_{0}^{\prime }$, using the QCD two-point sum rule method, and two
interpolating currents. The widths of $X_{0}$ and $X_{0}^{\prime }$ are
calculated in the framework of the light-cone sum rule (LCSR) method. This
is necessary to find strong couplings at vertices $X_{0}^{(\prime
)}D^{-}K^{+}$ and $X_{0}^{(\prime )}\overline{D}^{0}K^{0}$ which determine
partial widths of the decay channels $X_{0}^{(\prime )}\rightarrow
D^{-}K^{+} $ and $X_{0}^{(\prime )}\rightarrow \overline{D}^{0}K^{0}$.
Because aforementioned strong couplings correspond to tetraquark-meson-meson
type vertices, the LCSR method is supplied by technique of a soft-meson
approximation.

This work is organized in the following way: In Section \ref{sec:Masses}, we
calculate masses and couplings of the ground-state and radially excited
tetraquarks $X_{0}^{(\prime )}$. To this end, we use both the scalar-scalar
and axial-axial type interpolating currents. The sum rule computations are
carried out by including effects of vacuum condensates up to dimension $15$.
In Section \ref{sec:Decays}, we compute the strong couplings $g^{(\prime )}$
and $G^{(\prime )}$ that describe strong interaction of particles at the
vertices $X_{0}^{(\prime )}D^{-}K^{+}$ and $X_{0}^{(\prime )}\overline{D}%
^{0}K^{0}$. Here, we evaluate also partial widths of the decays $%
X_{0}^{(\prime )}\rightarrow D^{-}K^{+}$ and $X_{0}^{(\prime )}\rightarrow
\overline{D}^{0}K^{0}$, and find full widths of the tetraquarks $%
X_{0}^{(\prime )}$. Section \ref{sec:Disc} is devoted to discussions and
conclusions.


\section{The mass and current coupling of $1S$ and $2S$ tetraquarks $X_{0}$
and $X_{0}^{\prime }$}

\label{sec:Masses}
The mass and current coupling of tetraquarks $X_{0}$ and $X_{0}^{\prime }$
are among their important parameters. The masses of these states are
necessary to compare them with the LHCb data and fix whether one of these
particles may be interpreted as the resonance $X_{0}(2900)$. The
current couplings of $X_{0}$ and $X_{0}^{\prime }$ in conjunctions with
their masses are required to calculate partial widths of the decay channels $%
X_{0}^{(\prime )}\rightarrow D^{-}K^{+}$ and $X_{0}^{(\prime )}\rightarrow
\overline{D}^{0}K^{0}$, and hence to evaluate full width of these
tetraquarks.

We compute the mass and coupling of $X_{0}$ and $X_{0}^{\prime }$ in the
framework the QCD two-point sum rule method, which is one of effective
nonperturbative approaches in the high energy physics \cite%
{Shifman:1978bx,Shifman:1978by}. It rests on fundamental principles of QCD
and leads to reliable predictions using as input parameters only few
universal vacuum condensates. Remarkably, sum rules derived by means of this
method are applicable to investigate not only ordinary, but also multiquark
hadrons \cite{Chen:2016qju,Chen:2016spr,Albuquerque:2018jkn,Agaev:2020zad}.

We start our study from consideration of the following two-point correlation
function
\begin{equation}
\Pi (p)=i\int d^{4}xe^{ipx}\langle 0|\mathcal{T}\{J(x)J^{\dag
}(0)\}|0\rangle ,  \label{eq:CF1}
\end{equation}%
where $\mathcal{T}$ means the time-ordered product, and $J(x)$ is the
interpolating current for the tetraquarks $X_{0}$ and $X_{0}^{\prime }$. In
general, tetraquarks $X_{0}$ and $X_{0}^{\prime }$ with required quantum
numbers $J^{\mathrm{P}}=0^{+}$ can be built of different diquarks: It may be
composed of scalar diquark and antidiquark pair $u^{T}C\gamma _{5}d$ and $%
\overline{c}\gamma _{5}C\overline{s}^{T}$ or made of an axial-vector diquark
$u^{T}C\gamma _{\mu }d$ and an axial-vector antidiquark $\overline{c}\gamma
^{\mu }C\overline{s}^{T}$, where $C$ is the charge conjugation matrix.
Interpolating currents that correspond to these structures have the
following forms%
\begin{equation}
J_{\mathrm{S}}(x)=\epsilon \widetilde{\epsilon }[u_{b}^{T}(x)C\gamma
_{5}d_{c}(x)][\overline{c}_{d}(x)\gamma _{5}C\overline{s}_{e}^{T}(x)],
\label{eq:CR2}
\end{equation}%
and
\begin{equation}
J(x)=\epsilon \widetilde{\epsilon }[u_{b}^{T}(x)C\gamma _{\mu }d_{c}(x)][%
\overline{c}_{d}(x)\gamma ^{\mu }C\overline{s}_{e}^{T}(x)],  \label{eq:CR1}
\end{equation}%
where $\epsilon \widetilde{\epsilon }=\epsilon _{abc}\widetilde{\epsilon }%
_{ade}$, and $a$, $b$, $c$, $d$ and $e$ are color indices. In Eqs.\ (\ref%
{eq:CR2}) and (\ref{eq:CR1}) $c(x)$, $s(x)$, $u(x)$ and $d(x)$ are
corresponding quark fields. In what follows, we consider in a detailed
manner the interpolating current $J(x)$, and provide only final results
obtained while employing $J_{\mathrm{S}}(x)$.

To derive required sum rules, the correlation function $\Pi (p)$ has to be
expressed in terms of $X_{0}$ and $X_{0}^{\prime }$ tetraquarks' physical
parameters. The function $\Pi ^{\mathrm{Phys}}(p)$ obtained after relevant
manipulations constitutes the physical (phenomenological) side of the sum
rules. We analyze a ground-state and first radially excited particles,
therefore include contributions of these states to the correlation function
explicitly. As a result, we obtain
\begin{equation}
\Pi ^{\mathrm{Phys}}(p)=\frac{\langle 0|J|X_{0}\rangle \langle
X_{0}|J^{\dagger }|0\rangle }{m^{2}-p^{2}}+\frac{\langle 0|J|X_{0}^{\prime
}\rangle \langle X_{0}^{\prime }|J^{\dagger }|0\rangle }{m^{\prime 2}-p^{2}}%
\cdots,  \label{eq:Phys1}
\end{equation}%
where $m$ and $m^{\prime }$ are the masses of the tetraquarks $X_{0}$ and $%
X_{0}^{\prime }$. The formula (\ref{eq:Phys1}) is derived by saturating the
correlation function $\Pi (p)$ with a full set of scalar four-quark states
and performing integration over $x$ in Eq.\ (\ref{eq:CF1}). Dots in Eq.\ (%
\ref{eq:Phys1}) stand for effects of higher resonances and continuum states
in the $X_{0}$ channel.

Equation (\ref{eq:Phys1}) contains two simple-pole terms, which in the case
of multiquark hadrons have to be used with some caution. The reason is that
the physical side may contain also two-meson reducible contributions.
Indeed, the current $J(x)$ couples not only to the tetraquarks $X_{0}$ and $%
X_{0}^{\prime }$, but also interacts with conventional two-meson states \cite%
{Kondo:2004cr,Lee:2004xk}. These two-meson contributions modify a quark
propagator in Eq.\ (\ref{eq:Phys1})
\begin{equation}
\frac{1}{m^{2}-p^{2}}\rightarrow \frac{1}{m^{2}-p^{2}-i\sqrt{p^{2}}\Gamma (p)%
},  \label{eq:Mod}
\end{equation}%
where $\Gamma (p)$ is the finite width of the tetraquark generated by
two-meson effects. They should be subtracted from the sum rules, or taken
into account in parameters of the pole terms. For tetraquarks the second
method was applied in articles \cite%
{Wang:2015nwa,Agaev:2018vag,Sundu:2018nxt}, and it was demonstrated that
these contributions can be absorbed into the current coupling keeping, at
the same time, stable the mass of the tetraquark. Detailed analyses proved
that two-meson effects are small, and do not exceed theoretical errors of
the sum rule method itself \cite%
{Lee:2004xk,Wang:2015nwa,Agaev:2018vag,Sundu:2018nxt}. Therefore, the
physical side of the sum rules is written down above by applying the
zero-width single-pole approximation.

Using the matrix elements
\begin{equation}
\langle 0|J|X_{0}^{(\prime )}\rangle =f^{(\prime )}m^{(\prime )},
\label{eq:ME1}
\end{equation}%
it is possible to simplify the function $\Pi ^{\mathrm{Phys}}(p)$. Simple
operations lead for $\Pi ^{\mathrm{Phys}}(p)$ to the expression%
\begin{equation}
\Pi ^{\mathrm{Phys}}(p)=\frac{f^{2}m^{2}}{m^{2}-p^{2}}+\frac{f^{\prime
2}m^{\prime 2}}{m^{\prime 2}-p^{2}}\cdots.  \label{eq:Phen2}
\end{equation}%
The function $\Pi ^{\mathrm{Phys}}(p)$ has a simple Lorentz structure $\sim
I $, and, depending on a problem under consideration, one or a sum of two
terms may form the corresponding invariant amplitude $\Pi ^{\mathrm{Phys}%
}(p^{2})$.

The second component of the sum rules $\Pi ^{\mathrm{OPE}}(p)$, should be
computed in the operator product expansion ($\mathrm{OPE}$) with certain
accuracy. It can be found by employing the expression of the interpolating
current $J(x)$, and replacing contracted quark fields by relevant
propagators. After these operations, we obtain for $\Pi ^{\mathrm{OPE}}(p)$
\begin{eqnarray}
&&\Pi ^{\mathrm{OPE}}(p)=i\int d^{4}xe^{ipx}\epsilon \widetilde{\epsilon }%
\epsilon ^{\prime }\widetilde{\epsilon }^{\prime }\mathrm{Tr}\left[
S_{s}^{e^{\prime }e}(-x)\gamma ^{\mu }\right.  \notag \\
&&\left. \times \widetilde{S}_{c}^{d^{\prime }d}(-x)\gamma ^{\nu }\right]
\mathrm{Tr}\left[ S_{u}^{bb^{\prime }}(x)\gamma _{\nu }\widetilde{S}%
_{d}^{cc^{\prime }}(x)\gamma _{\mu }\right] ,  \label{eq:QCD1}
\end{eqnarray}%
where%
\begin{equation}
\widetilde{S}_{c(q)}(x)=CS_{c(q)}^{T}(x)C.
\end{equation}%
Here, $S_{c}(x)$ and $S_{q}(x)$ are the heavy $c$- and light $q=u(s,d)$%
-quark propagators, respectively. Their explicit expressions are collected
in Appendix. The correlation function $\Pi ^{\mathrm{OPE}}(p)$ has a simple
Lorentz structure: We use for a corresponding invariant amplitude a notation
$\Pi ^{\mathrm{OPE}}(p^{2})$.

The correlation function $\Pi ^{\mathrm{Phys}}(p)$ corresponds to the
"ground-state+excited particle+continuum" scheme, and encompasses
contributions of two particles. At the first stage of studies, we employ a
familiar "ground-state+continuum" scheme, and find the mass and coupling of
the ground-state tetraquark $X_{0}$. This means that, we include the second
term in $\Pi ^{\mathrm{Phys}}(p)$ into a list of "higher resonances and
continuum states", and get the standard expression for the correlation
function. Following operations are well known, and were discussed repeatedly
in the literature including our papers. Therefore, we skip further details
and provide final formulas for $m$ and $f$ :
\begin{equation}
m^{2}=\frac{\Pi ^{\prime }(M^{2},s_{0})}{\Pi (M^{2},s_{0})},
\label{eq:Mass1}
\end{equation}%
and
\begin{equation}
f^{2}=\frac{e^{m^{2}/M^{2}}\Pi (M^{2},s_{0})}{m^{2}},  \label{eq:CP1}
\end{equation}%
where $M^{2}$ and $s_{0}$ are the Borel and continuum threshold parameters,
respectively. Here, $\Pi (M^{2},s_{0})$ is the Borel transformed and
subtracted invariant amplitude $\Pi ^{\mathrm{OPE}}(p^{2})$, and $\Pi
^{\prime }(M^{2},s_{0})=d\Pi (M^{2},s_{0})/d(-1/M^{2})$.

At this stage, one should fix the working windows for the parameters $M^{2}$
and $s_{0}$, which are auxiliary quantities of sum rule computations and
should obey some important restrictions. The dominance the pole contribution
($\mathrm{PC}$), convergence of $\mathrm{OPE,}$ and stability of physical
quantities against variations of the Borel parameter are main constraints
imposed on the correlation function $\Pi (M^{2},s_{0})$. Fulfilment of these
constraints can be established using expressions
\begin{equation}
\mathrm{PC}=\frac{\Pi (M^{2},s_{0})}{\Pi (M^{2},\infty )},  \label{eq:PC}
\end{equation}%
and
\begin{equation}
R(M^{2})=\frac{\Pi ^{\mathrm{DimN}}(M^{2},s_{0})}{\Pi (M^{2},s_{0})},
\label{eq:Convergence}
\end{equation}%
and numerical limits on $\mathrm{PC}$, $R(M^{2})$, as well as fixing
acceptable variations of $m$ and $f$. Let us note, that in Eq.\ (\ref%
{eq:Convergence}) $\Pi ^{\mathrm{DimN}}(M^{2},s_{0})$ is a last term or a
sum of last few terms in the correlation function. In the present paper, we
employ last three terms in the $\mathrm{OPE}$, and hence $\Pi ^{\mathrm{DimN}%
}(M^{2},s_{0})=\Pi ^{\mathrm{Dim(13+14+15)}}(M^{2},s_{0})$.

Having fixed working regions for $M^{2}$ and $s_{0}$, one can extract the
mass and coupling of the $1S$ tetraquark $X_{0}$. The quantites $m$ and $f$,
strictly speaking, should not depend on the Borel parameter. But there are
residual effects of working regions on extracted parameters, which
nevertheless have to stay within acceptable limits. On the contrary, the
continuum threshold parameter $s_{0}$ bears physical information about the
mass of the excited tetraquark $X_{0}^{\prime }$. In fact, the parameter $%
s_{0}$ separates contribution of the ground-state particle from ones due to
higher resonances and continuum states. This means, that masses of $X_{0}$
and $X_{0}^{\prime }$ must obey restrictions $m<\sqrt{s_{0}}\leq m^{\prime }$%
.

After calculating the mass and coupling of the $X_{0}$, we can find
parameters of the excited state $X_{0}^{\prime }$. For these purposes, we
treat $m$ and $f$ as input parameters and look for new working regions for $%
M^{2}$ and $s_{0}^{\ast }$ which have to satisfy not only Eqs.\ (\ref{eq:PC}%
) and (\ref{eq:Convergence}), but also obey $s_{0}^{\ast }>s_{0}$. Necessity
of last constraint is evident, because in the "ground-state+excited
particle+continuum" scheme the parameter $s_{0}^{\ast }$ separates two
states from remaining higher resonances. The mass of the $X_{0}^{\prime }$
extracted from a new sum rule is bounded by conditions $\sqrt{s_{0}}\leq
m^{\prime }<\sqrt{s_{0}^{\ast }}$. If regions for $M^{2}$ and $s_{0}^{\ast }$%
, and extracted mass $m^{\prime }$ comply with these regulations, performed
analysis is self-consistent and gives reliable predictions.

The sum rules for $m^{\prime }$ and $f^{\prime }$ obviously differ from ones
for $m$ and $f$. For the mass $m^{\prime }$, we derive the following
expression%
\begin{equation}
m^{\prime 2}=\frac{\Pi ^{\prime }(M^{2},s_{0}^{\ast
})-f^{2}m^{4}e^{-m^{2}/M^{2}}}{\Pi (M^{2},s_{0}^{\ast
})-f^{2}m^{2}e^{-m^{2}/M^{2}}},  \label{eq:Mass2}
\end{equation}%
whereas for $f^{\prime }$ get%
\begin{equation}
f^{\prime 2}=\frac{e^{m^{\prime 2}/M^{2}}\left[ \Pi (M^{2},s_{0}^{\ast
})-f^{2}m^{2}e^{-m^{2}/M^{2}}\right] }{m^{\prime 2}}.  \label{eq:CP2}
\end{equation}%
It is evident, that parameters $m^{\prime }$ and $f^{\prime }$ of the
excited particle $X_{0}^{\prime }$ depend explicitly on the mass and current
coupling of the ground-state tetraquark $X_{0}$. Such dependence is natural,
because Eq.\ (\ref{eq:Phen2}) contains two terms, and $m$ and $f$ appear as
inputs when calculating $m^{\prime }$ and $f^{\prime }$. In its turn, the
excited state $X_{0}^{\prime }$ also affects the mass $m$ and coupling\ $f$
of the ground-state particle, but its effect is implicit and encoded in a
choice of the continuum threshold parameter $s_{0}$. In fact, the parameters
$m$ and \ $f$, extracted from sum rules depend on the correlation function $%
\Pi (M^{2},s_{0})$ at $s_{0}$, which is limited by the mass $m^{\prime }$ of
the excited state $\sqrt{s_{0}}\leq m^{\prime }$. Because two sets $(m,\ f)$
and $(m^{\prime },\ f")$ are determined by the same correlation function at
different $s_{0}$ and $s_{0}^{\ast }$, one may consider a difference of $\Pi
(M^{2},s_{0})$ at $s_{0}$ and $s_{0}^{\ast }$ as a "measure" of this effect.

The correlation function $\Pi (M^{2},s_{0})$ has the following form%
\begin{equation}
\Pi (M^{2},s_{0})=\int_{\mathcal{M}^{2}}^{s_{0}}ds\rho ^{\mathrm{OPE}%
}(s)e^{-s/M^{2}}+\Pi (M^{2}),  \label{eq:InvAmp}
\end{equation}%
where $\mathcal{M}=m_{c}+m_{s}$. In current work, we neglect masses of the
quarks $u$ and $d$, and terms $\sim m_{s}^{2}$, but take into account
contributions of $m_{s}$. The spectral density $\rho ^{\mathrm{OPE}}(s)$ is
calculated as an imaginary part of the correlator $\Pi ^{\mathrm{OPE}}(p)$.
The function $\Pi (M^{2})$ is the Borel transformation of terms in $\Pi ^{%
\mathrm{OPE}}(p)$ derived directly from their expressions. Computations are
performed by including into analysis vacuum condensates till dimension $15$.
In Appendix, for the sake of brevity,\ we provide analytical expressions of $%
\rho ^{\mathrm{OPE}}(s)$ and $\Pi (M^{2})$ up to dimension $11$.

Our analytical results contain nonperturbative terms up to dimension $15$,
which makes necessary to explain treatment of higher dimensional vacuum
condensates. The propagator $S_{q}(x)$ contains various quark, gluon and
mixed condensates of different dimensions, terms proportional to $%
g_{s}^{2}G^{2}$ and $g_{s}^{3}G^{3}$ are taken into account in $S_{c}(x)$.
Some of terms in the propagator $S_{q}(x)$, for instance, ones proportional
to $\langle \overline{q}g_{s}\sigma Gq\rangle $, $\langle \overline{q}%
q\rangle ^{2}$, and $\langle \overline{q}q\rangle \langle
g_{s}^{2}G^{2}\rangle $ are obtained using the factorization hypothesis of
higher dimensional condensates. These terms and their products with
condensates from other light quark propagators, as well as with relevant
components of $S_{c}(x)$ enter to $\rho ^{\mathrm{OPE}}(s)$ and $\Pi (M^{2})$%
. We carry out computations by taking into account all contributions up to
dimension $15$ obtained by this way. But factorization of higher dimensional
condensates is not precise and generates uncertainties \cite{Ioffe:2005ym},
which sometimes are difficult to estimate. Because contributions of higher
dimensional terms are numerically very small, we neglect impact of such
uncertainties on extracted quantities.

The sum rules for $m^{(\prime )}$ and $f^{(\prime )}$ contain universal
quark, gluon and mixed vacuum condensates listed below
\begin{eqnarray}
&&\langle \overline{q}q\rangle =-(0.24\pm 0.01)^{3}~\mathrm{GeV}^{3},\
\langle \overline{s}s\rangle =(0.8\pm 0.1)\langle \overline{q}q\rangle ,
\notag \\
&&\langle \overline{q}g_{s}\sigma Gq\rangle =m_{0}^{2}\langle \overline{q}%
q\rangle ,\ \langle \overline{s}g_{s}\sigma Gs\rangle =m_{0}^{2}\langle
\overline{s}s\rangle ,  \notag \\
&&m_{0}^{2}=(0.8\pm 0.2)~\mathrm{GeV}^{2}  \notag \\
&&\langle \frac{\alpha _{s}G^{2}}{\pi }\rangle =(0.012\pm 0.004)~\mathrm{GeV}%
^{4},  \notag \\
&&m_{s}=93_{-5}^{+11}~\mathrm{MeV},\ m_{c}=1.27\pm 0.02~\mathrm{GeV}.
\label{eq:Parameters}
\end{eqnarray}%
The masses of $c$ and $s$ quarks are also included into Eq.\ (\ref%
{eq:Parameters}).

\begin{widetext}

\begin{figure}[h!]
\begin{center}
\includegraphics[totalheight=6cm,width=8cm]{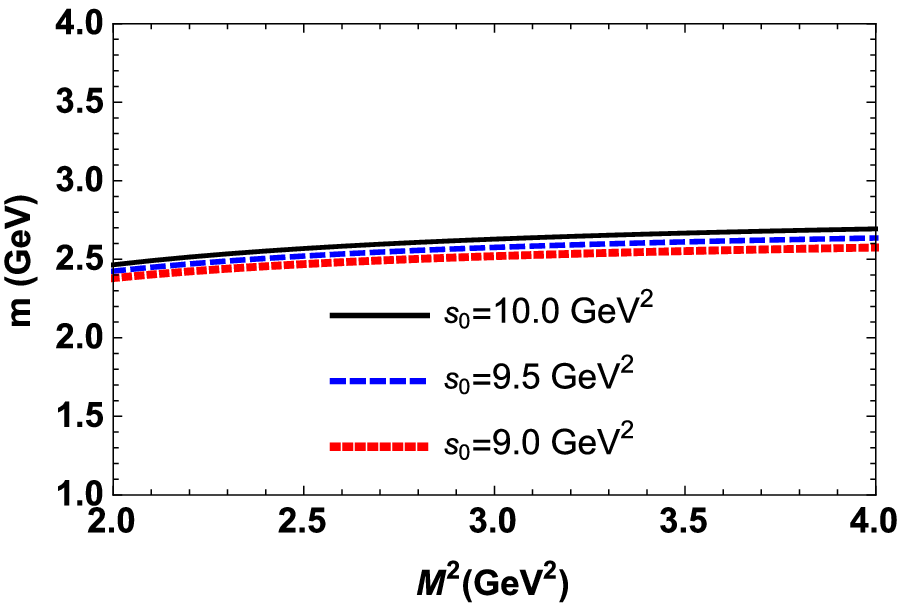}
\includegraphics[totalheight=6cm,width=8cm]{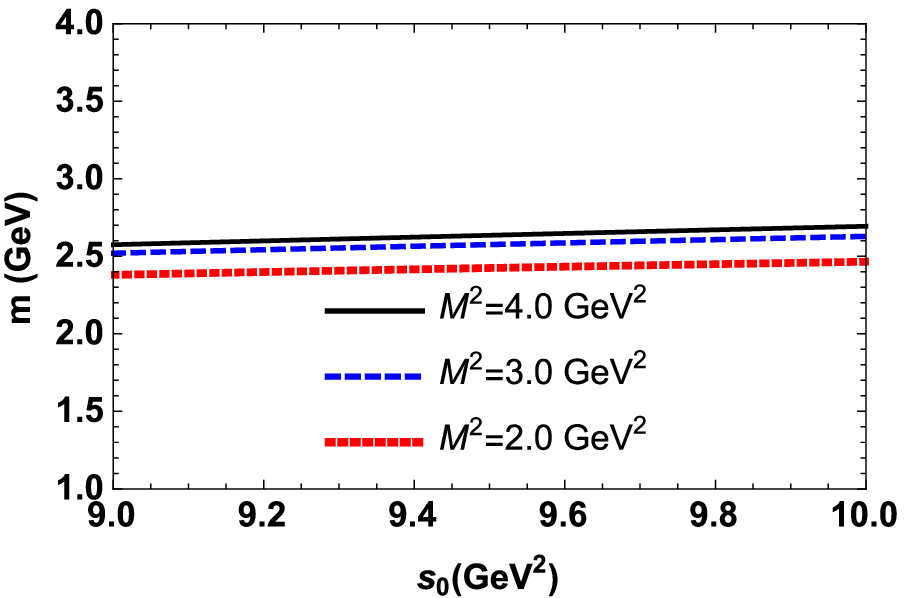}
\end{center}
\caption{The mass $m$ of the tetraquark $X_0$ as a function of the Borel parameter $M^{2}$ [left panel], and as a function of $s_0$ [right panel].}
\label{fig:Mass}
\end{figure}
\begin{figure}[h!]
\begin{center}
\includegraphics[totalheight=6cm,width=8cm]{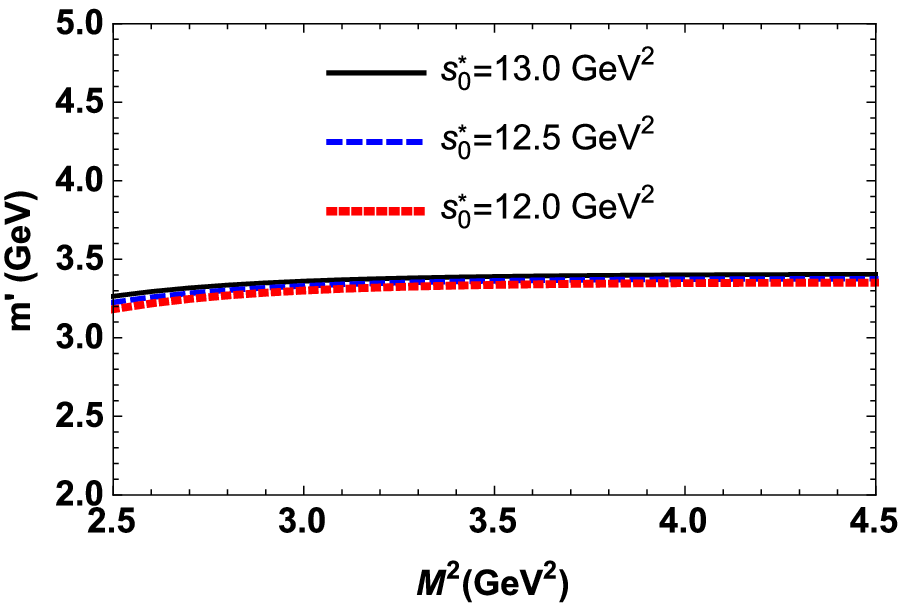}
\includegraphics[totalheight=6cm,width=8cm]{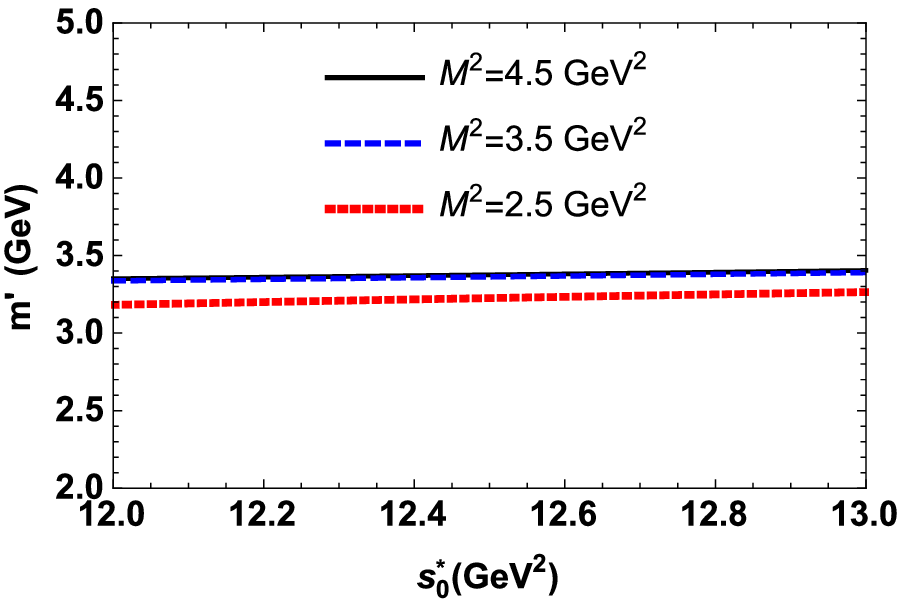}
\end{center}
\caption{The same as in Fig.\ 1, but for the mass $m^{\prime}$ of the excited tetraquark $X_0^{\prime}$.}
\label{fig:MassS0}
\end{figure}

\end{widetext}

We begin from analysis of the ground-state tetraquark $X_{0}$, and fix
regions $M^{2}$ and $s_{0}$, in which its parameters can be extracted. We
determine the region for $M^{2}$ by requiring fulfilment of the condition $%
\mathrm{PC}\geq 0.2$ at maximal value of $M_{\mathrm{max}}^{2}$ and
convergence of $\mathrm{OPE}$ at its minimum, i.e., $R(M_{\mathrm{min}%
}^{2})\leq 0.01$. Our calculations demonstrate, that the working regions
\begin{equation}
M^{2}\in \lbrack 2,4]\ \mathrm{GeV}^{2},\ s_{0}\in \lbrack 9,10]~\mathrm{GeV}%
^{2},  \label{eq:Wind1}
\end{equation}%
satisfy aforementioned restrictions. Thus, at $M_{\mathrm{max}}^{2}=4~%
\mathrm{GeV}^{2}$ the pole contribution is equal to $0.23$, whereas at $M_{%
\mathrm{min}}^{2}=2~\mathrm{GeV}^{2}$ it equals to $0.7$. At $M_{\mathrm{min}%
}^{2}=2~\mathrm{GeV}^{2}$, we get $R(M_{\mathrm{min}}^{2})<0.01$, hence the
convergence of the sum rules is ensured. Mean values of $m$ and $f$ averaged
over the regions (\ref{eq:Wind1}) read
\begin{eqnarray}
m &=&(2545~\pm 160)~\mathrm{MeV},  \notag \\
f &=&(3.0\pm 0.5)\times 10^{-3}~\mathrm{GeV}^{4}.  \label{eq:Result1}
\end{eqnarray}

Uncertainties of the results in Eq.\ (\ref{eq:Result1}) are within
acceptable limits: for the mass and coupling they form $\pm 6.3\%$ and $\pm
16.7\%$ of the corresponding central values, respectively. Theoretical
uncertainties of $m$ are smaller, because the relevant sum rule Eq.\ (\ref%
{eq:Mass1}) is given as a ratio of correlation functions, whereas $f$ \ is
determined by the expression with the correlation function in the numerator
of Eq.\ (\ref{eq:CP1}). On the Fig.\ \ref{fig:Mass}, we depict the sum
rule's prediction for $m$ as functions of $M^{2}$ and $s_{0}$ in which one
can look at dependence of $m$ on the Borel and continuum threshold
parameters.

\begin{table}[tbp]
\begin{tabular}{|c|c|c|}
\hline\hline
Tetraquarks & $X_{\mathrm{S}}$ & $X_{\mathrm{S}}^{\prime}$ \\ \hline\hline
$M^2 ~(\mathrm{GeV}^2$) & $2-4$ & $2.5-4.5$ \\ \hline
$s_0(s_0^{\star}) ~(\mathrm{GeV}^2$) & $9-10$ & $12-13$ \\ \hline
$m_{\mathrm{S}} ~(\mathrm{MeV})$ & $2663 \pm 110 $ & $3325 \pm 85 $ \\ \hline
$f_{\mathrm{S}} \cdot 10^{3} ~(\mathrm{GeV}^4)$ & $2.2 \pm 0.3$ & $2.7 \pm 0.4$ \\
\hline\hline
\end{tabular}%
\caption{The mass and current coupling of the tetraquarks $X_{\mathrm{S}}$
and $X_{\mathrm{S}}^{\prime}$, and parameters $M^2$ and $s_0$ used in their
computations.}
\label{tab:Results1}
\end{table}

To find parameters of the first radially excited tetraquark $X_{0}^{\prime }$%
, we start our analysis from Eqs.\ (\ref{eq:Mass2}) and (\ref{eq:CP2}) and
explore regions for $M^{2}$ and $s_{0}^{\ast }$ bearing in mind that $%
s_{0}^{\ast }>s_{0}$. It is not difficult to see that working windows%
\begin{equation}
M^{2}\in \lbrack 2.5,4.5]~\mathrm{GeV}^{2},\ s_{0}^{\ast }\in \lbrack 12,13]~%
\mathrm{GeV}^{2},  \label{eq:Wind2}
\end{equation}%
obey necessary constraints. In these regions the pole contribution to $\Pi
(M^{2},s_{0}^{\ast })$ changes inside of \ interval%
\begin{equation}
0.75\geq \mathrm{PC}\geq 0.34.
\end{equation}%
The mass and coupling of the radially excited tetraquark are
\begin{eqnarray}
m^{\prime } &=&(3320~\pm 120)~\mathrm{MeV},  \notag \\
f^{\prime } &=&(3.7\pm 0.6)\times 10^{-3}~\mathrm{GeV}^{4},
\label{eq:Result2}
\end{eqnarray}%
respectively. Dependence of $m^{\prime }$ on the parameters $M^{2}$ and $%
s_{0}^{\ast }$ is shown on Fig.\ \ref{fig:MassS0}. Comparing figures \ref%
{fig:Mass} and \ref{fig:MassS0} one sees, that theoretical ambiguities for
the mass of the tetraquark $X_{0}^{\prime }$ are smaller than that for $m$.

With these final predictions in hand, one can check self-consistency of
performed analysis. Using mean values of the parameters $\sqrt{s_{0}^{\ast }}%
=3.54~\mathrm{GeV}$ and $\sqrt{s_{0}}=3.08~\mathrm{GeV}$ it is easy to be
convinced that all regulations discussed above are correct.

The mass and coupling of the ground-state and excited tetraquarks $X_{%
\mathrm{S}}$ and $X_{\mathrm{S}}^{\prime }$ extracted from the sum rules by
employing the interpolating current $J_{\mathrm{S}}(x)$ are shown in Table\ %
\ref{tab:Results1}. We plot also the masses $m_{\mathrm{S}}$ and $m_{\mathrm{%
S}}^{\prime }$ in Figs.\ \ref{fig:Mass1} and \ref{fig:Mass1S0} as functions
of the Borel and continuum threshold parameters.

\begin{widetext}

\begin{figure}[h!]
\begin{center}
\includegraphics[totalheight=6cm,width=8cm]{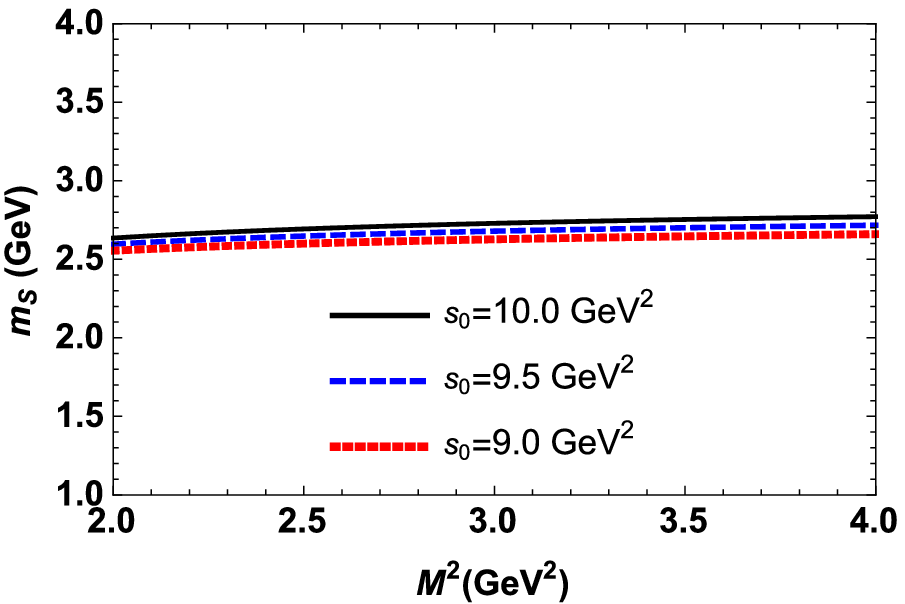}
\includegraphics[totalheight=6cm,width=8cm]{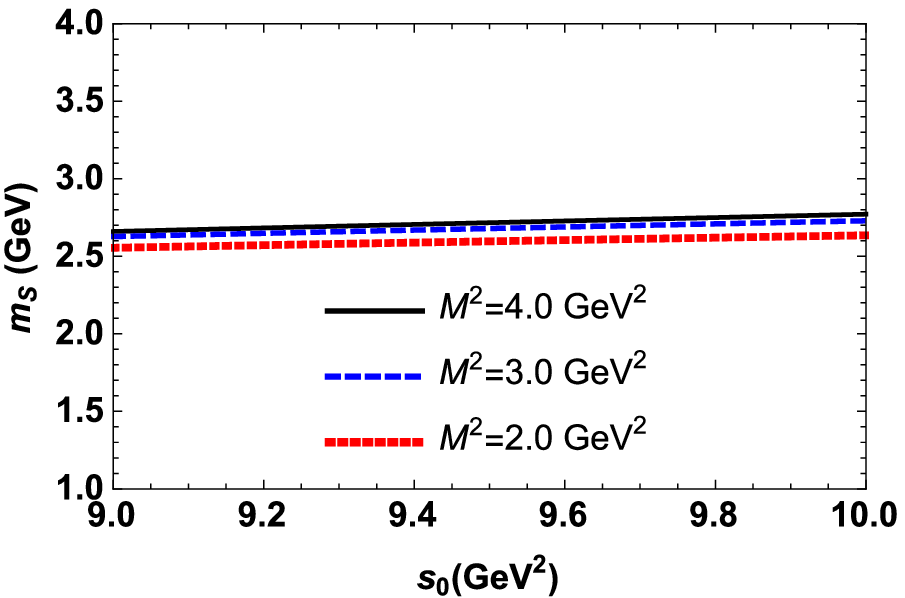}
\end{center}
\caption{Dependence of the mass $m_{\mathrm{S}}$  on the
Borel parameter $M^{2}$ at some fixed $s_{0}$ [left panel] and on
the continuum
threshold parameter $s_{0}$ at fixed Borel parameter
[right panel].}
\label{fig:Mass1}
\end{figure}
\begin{figure}[h!]
\begin{center}
\includegraphics[totalheight=6cm,width=8cm]{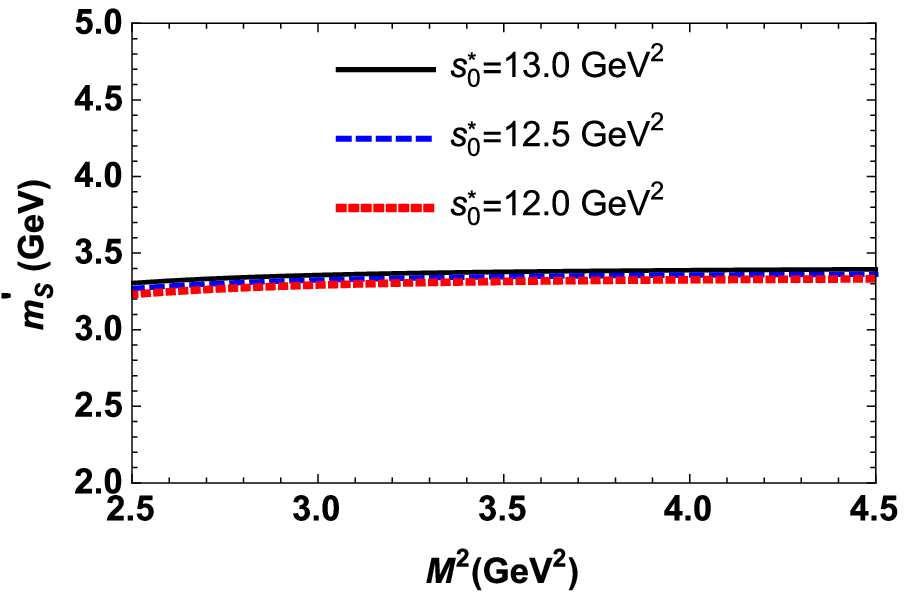}%
\includegraphics[totalheight=6cm,width=8cm]{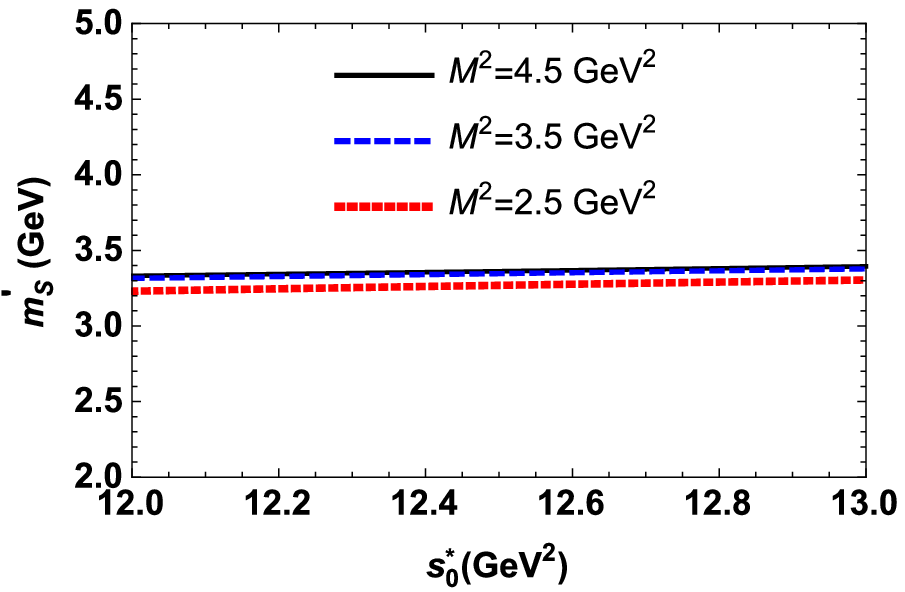}
\end{center}
\caption{The same as in Fig.\ 3, but for the mass $m_{\mathrm{S}}%
^{\prime}$ of the excited state.}
\label{fig:Mass1S0}
\end{figure}

\end{widetext}

Results obtained for the masses of the states $X_{0}$ and $X_{0}^{\prime }$
are either smaller than the LHCb data for the resonance $X_{0}(2900)$, as in
the case of the ground-state tetraquark $X_{0}$, or exceed it. These
conclusions are valid for both currents $J(x)$ and $J_{\mathrm{S}}(x)$, and
even ambiguities of calculations, taken into account in $m^{(\prime )}$ and $%
m_{\mathrm{S}}^{(\prime )}$, do not solve the problem. It seems, that the
diquark-antidiquark structure of $X_{0}$ and its radial excitation $%
X_{0}^{\prime }$ are new exotic mesons not yet seen in experiments. To gain
detailed information on their properties, we consider decays of the
tetraquarks $X_{0}$ and $X_{0}^{\prime }$, and estimate their full widths in
the next section.

\section{Processes $X_{0}^{(\prime )}\rightarrow D^{-}K^{+}$ and $%
X_{0}^{(\prime )}\rightarrow \overline{D}^{0}K^{0}$}

\label{sec:Decays}

Masses of the tetraquarks $X_{0}^{(\prime )}$ calculated in the previous
section, as well as their quark content allow us to specify their decay
channels. It is not difficult to see that thresholds $\approx 2364~\mathrm{%
MeV}$ for production of conventional meson pairs $D^{-}K^{+}$ and $\overline{%
D}^{0}K^{0}$ are smaller than masses of $X_{0}^{(\prime )}$. Moreover the
modes $X_{0}^{(\prime )}\rightarrow D^{-}K^{+}$ and $X_{0}^{(\prime
)}\rightarrow \overline{D}^{0}K^{0}$ are $S$-wave decay channels for the
tetraquarks $X_{0}^{(\prime )}$, and decay to mesons $D^{-}K^{+}$ is
dominant process for the resonance $X_{0}(2900)$.

In this section, we consider in a rather detailed form decays $%
X_{0}^{(\prime )}\rightarrow D^{-}K^{+}$, and provide final information
about channels $X_{0}^{(\prime )}\rightarrow \overline{D}^{0}K^{0}$. Partial
widths of the processes $X_{0}\rightarrow D^{-}K^{+}$ and $X_{0}^{\prime
}\rightarrow D^{-}K^{+}$are determined by strong couplings at corresponding
tetraquark-meson-meson vertices $X_{0}D^{-}K^{+}$ and $X_{0}^{\prime
}D^{-}K^{+}$, respectively. We denote relevant strong couplings as $g$ and $%
g^{\prime }$, and use for their calculations the QCD sum rules on the
light-cone \cite{Balitsky:1989ry,Belyaev:1994zk}, and techniques of the
soft-meson approximation \cite{Ioffe:1983ju}.

The strong couplings $g$ and $g^{\prime }$ are defined by the on-mass-shell
matrix element%
\begin{equation}
\langle K(q)D\left( p\right) |X_{0}^{(\prime )}(p^{\prime })\rangle
=g^{(\prime )}p\cdot p^{\prime }.  \label{eq:Mel1}
\end{equation}%
In the framework of the LCSR method the vertex $X_{0}D^{-}K^{+}$ can be
investigated by means of the correlation function
\begin{equation}
\Pi (p,q)=i\int d^{4}xe^{ipx}\langle K(q)|\mathcal{T}\{J^{D}(x)J^{\dag
}(0)\}|0\rangle ,  \label{eq:CorrF3}
\end{equation}%
where mesons $K^{+}$ and $D^{-}$ are shortly denoted by $K$ and $D$,
respectively. In Eq.\ (\ref{eq:CorrF3}) $J(x)$ and $J^{D}(x)$ are the
interpolating currents for the tetraquarks $X_{0}^{(\prime )}$ and meson $%
D^{-}$. First of them is defined by Eq.\ (\ref{eq:CR1}), and for $J^{D}(x)$,
we employ
\begin{equation}
J^{D}(x)=\overline{c}_{j}(x)i\gamma _{5}d_{j}(x),  \label{eq:Dcur}
\end{equation}%
with $j$ being the color index.

The current $J(x)$ couples to both the ground-state and radially excited
tetraquarks $X_{0}$ and $X_{0}^{\prime }$, therefore in the function $\Pi ^{%
\mathrm{Phys}}(p,q)$ we should take into account contribution of these
particles explicitly. We are interested in terms which have poles at
variables $p^{2}$ and $p^{\prime 2}$, where $p$ and $p^{\prime }=p+q$ are
the momenta of the meson $D^{-}$ and tetraquarks $X_{0}^{(\prime )}$, and $q$
is momentum of the $K^{+}$ meson. The terms in $\Pi ^{\mathrm{Phys}}(p,q)$
necessary for our analysis have the following forms%
\begin{eqnarray}
\Pi ^{\mathrm{Phys}}(p,q) &=&\frac{f_{D}m_{D}^{2}}{m_{c}(p^{2}-m_{D}^{2})}%
\left[ \frac{gfm}{(p^{\prime 2}-m^{2})}+\frac{g^{\prime }f^{\prime
}m^{\prime }}{(p^{\prime 2}-m^{\prime 2})}\right]  \notag \\
&&\times p\cdot p^{\prime }+\cdots ,  \label{eq:CorrF4a}
\end{eqnarray}%
where $m_{D}$ and $f_{D}$ are the mass and decay constant of the $D^{-}$
meson. To derive Eq.\ (\ref{eq:CorrF4a}) we use the vertex function given by
Eq.\ (\ref{eq:Mel1}), well known matrix elements of the tetraquarks $%
X_{0}^{(\prime )}$ Eq.\ (\ref{eq:ME1}), and new matrix element of the $D^{-}$
meson
\begin{equation}
\langle 0|J^{D}|D\left( p\right) \rangle =\frac{f_{D}m_{D}^{2}}{m_{c}}.
\label{eq:Mel}
\end{equation}%
The terms presented explicitly in Eq. (\ref{eq:CorrF4a}) correspond to
ground-state meson in $D^{-}$ channel, and ground-state and radially excites
tetraquarks in $X_{0}$ channel. Contributions of remaining higher resonances
and continuum states in the $D^{-}$ and $X_{0}$ channels are denoted by dots.

An expression of the same correlation function obtained using quark-gluon
degrees of freedom forms the second component $\Pi ^{\mathrm{QCD}}(p,q)$ of
the sum rule analysis. Calculations carried out using quark propagators give
\begin{eqnarray}
&&\Pi ^{\mathrm{OPE}}(p,q)=\int d^{4}xe^{ipx}\epsilon \widetilde{\epsilon }%
\left[ \gamma ^{\mu }\widetilde{S}_{d}^{jc}(x){}\gamma _{5}\right.  \notag \\
&&\left. \times \widetilde{S}_{c}^{jd}(-x){}\gamma _{\mu }\right] _{\alpha
\beta }\langle K(q)|\overline{u}_{\alpha }^{b}(0)s_{\beta }^{e}(0)|0\rangle ,
\label{eq:CorrF6}
\end{eqnarray}%
with $\alpha $ and $\beta $ being the spinor indices. The correlator $\Pi ^{%
\mathrm{OPE}}(p,q)$ contains quark propagators, which determine a hard-part
of this function. But it depends also on $\overline{u}s$ operator's local
matrix elements: this is soft factor in $\Pi ^{\mathrm{OPE}}(p,q)$.

The matrix elements $\langle K|\overline{u}s|0\rangle $ bear spinor and
color indices, and are inconvenient for further usage. To recast them into
color-singlet form and factor out spinor indices, we expand $\overline{u}s$
over the full set of Dirac matrices $\Gamma ^{J}$
\begin{equation}
\Gamma ^{J}=\mathbf{1},\ \gamma _{5},\ \gamma _{\mu },\ i\gamma _{5}\gamma
_{\mu },\ \sigma _{\mu \nu }/\sqrt{2},  \label{eq:Dirac}
\end{equation}%
and project them onto the colorless states
\begin{equation}
\overline{u}_{\alpha }^{b}(0)s_{\beta }^{a}(0)\rightarrow \frac{1}{12}\delta
^{ba}\Gamma _{\beta \alpha }^{J}\left[ \overline{u}(0)\Gamma ^{J}s(0)\right]
.  \label{eq:MatEx}
\end{equation}%
Obtained operators placed between the $K$ meson and vacuum give rise to
local matrix elements of the $K$ meson.

When considering the tetraquark-meson-meson vertices $X_{0}^{(\prime
)}D^{-}K^{+}$, we encounter the correlation function containing only local
matrix elements of quark operators. Let us note that such behavior of $\Pi ^{%
\mathrm{OPE}}(p,q)$ is typical for all vertices built of one tetraquark and
two conventional mesons. The reason is actually very simple: The tetraquark
current $J(0)$ is composed of four quark fields at the same space-time
position. Contractions of relevant fields from interpolating currents $%
J^{D}(x)$ and $J^{\dagger }(0)$ leave two free quark fields at the
space-time point $x=0$. As a result, local matrix elements of the $K$ meson
appear in the correlation function as overall normalization factors.

It is instructive to compare this situation with three-meson vertices, in
which contractions of quark fields with different space-time coordinates
generate $\Pi ^{\mathrm{OPE}}(p,q)$ containing non-local operators. Then
manipulations performed in accordance with Eqs. (\ref{eq:Dirac}) and (\ref%
{eq:MatEx}) lead to operators, matrix elements of which are distribution
amplitudes (DAs) of a final-state meson. In other words, for a three-meson
vertex a correlation function depends on integrals over DAs of a meson. A
situation described above in the LCSR method emerges in the kinematical
limit $q\rightarrow 0$ known as a soft-meson approximation \cite%
{Belyaev:1994zk}. In this approximation instead of a light-cone expansion,
one gets expansion in terms of local matrix elements of a final meson.
Because in the soft limit phenomenological and QCD sides of the light-cone
sum rules acquire distinctive features, they have to be treated in
accordance with elaborated methods \cite{Belyaev:1994zk,Ioffe:1983ju}. It is
important that strong couplings at three-meson vertices calculated using the
full version of the LCSR method and soft-meson approximation lead to
predictions, which are numerically very close to each other \cite%
{Belyaev:1994zk}.

The soft-meson approximation were applied to explore tetraquark-meson-meson
vertices in Ref.\ \cite{Agaev:2016dev}, and used later in numerous similar
studies \cite{Agaev:2020zad}. It is worth emphasizing, that in exclusive
processes with two tetraquarks and an ordinary meson correlation functions
contain integrals over DAs of a meson, and their treatment does not differ
from standard LCSR analysis \cite{Agaev:2016srl}.

Here, we employ this technique to analyze the vertices $X_{0}^{(\prime
)}\rightarrow D^{-}K^{+}$. \ As is seen from Eq.\ (\ref{eq:CorrF6}), the
soft-meson approximation considerably simplifies the QCD side of sum rules:
There are only local matrix elements of the $K$ meson in $\Pi ^{\mathrm{OPE}%
}(p^{2})$, and only a few of them contribute at the limit $q=0$. On the
contrary, the physical side of the sum rule has more complicated structure
than in the case of the full version of the LCSR method. The soft limit
implies fulfilment of the equality $p=p^{\prime }$, hence in the limit $%
q\rightarrow 0$ invariant amplitudes $\Pi ^{\mathrm{Phys}}(p^{2},p^{\prime
2})$ and $\Pi ^{\mathrm{OPE}}(p^{2},p^{\prime 2})$ are functions of a
variable $p^{2}$. Therefore, in Eq.\ (\ref{eq:CorrF4a}) one should take into
account that $p^{2}=p^{\prime 2}$, and gets
\begin{eqnarray}
\Pi ^{\mathrm{Phys}}(p^{2}) &=&\frac{f_{D}m_{D}^{2}}{m_{c}}\left[ gfm\frac{%
\widetilde{m}^{2}}{\left( p^{2}-\widetilde{m}^{2}\right) ^{2}}\right.  \notag
\\
&&\left. +g^{\prime }f^{\prime }m^{\prime }\frac{\widetilde{m}^{\prime 2}}{%
\left( p^{2}-\widetilde{m}^{\prime 2}\right) ^{2}}\right] +\cdots ,
\label{eq:CF2}
\end{eqnarray}%
where $\widetilde{m}^{2}=(m^{2}+m_{D}^{2})/2$ and $\widetilde{m}^{\prime
2}=(m^{\prime 2}+m_{D}^{2})/2$, respectively. Remaining problems are
connected with the Borel transform of the amplitude $\Pi ^{\mathrm{Phys}%
}(p^{2})$, which due to double poles at $p^{2}=\widetilde{m}^{2}$ and $p^{2}=%
\widetilde{m}^{\prime 2}$ has the following form
\begin{eqnarray}
\Pi ^{\mathrm{Phys}}(p^{2}) &=&\frac{f_{D}m_{D}^{2}}{m_{c}}\left[ gfm\frac{%
\widetilde{m}^{2}e^{-\widetilde{m}^{2}/M^{2}}}{M^{2}}\right. ,  \notag \\
&&\left. +g^{\prime }f^{\prime }m^{\prime }\frac{\widetilde{m}^{\prime 2}e^{-%
\widetilde{m}^{\prime 2}/M^{2}}}{M^{2}}\right] +\cdots .  \label{eq:CF3}
\end{eqnarray}

In general, the Borel transformation applied to a correlation function
suppresses contributions of higher resonances and continuum states. This
allows one to subtract these terms from the QCD side of the sum rule using
an assumption about quark-hadron duality. In the soft approximation, after
the Borel transformation there are still unsuppressed terms in the physical
side of the sum rule, which contribute to $\Pi ^{\mathrm{Phys}}(p^{2})$ on
an equal footing with ground-state term. Because we are interested in
analysis of both the ground-state $X_{0}$ and excited $X_{0}^{\prime }$
particles, it is necessary to clarify a nature of these unsuppressed terms.
The main contribution to $\Pi ^{\mathrm{Phys}}(p^{2})$ comes from the vertex
$X_{0}D^{-}K^{+}$, where the tetraquark and mesons are ground-state
particles. Unsuppressed terms correspond to vertices, in which $X_{0}$ is on
its excited state. While considering the vertex $X_{0}D^{-}K^{+}$ such
contributions should be treated as contaminations and removed applying some
procedures. Such prescriptions are well known and were described in Refs.\
\cite{Belyaev:1994zk,Ioffe:1983ju}: To eliminate contaminations from $\Pi ^{%
\mathrm{Phys}}(p^{2})$, one has to apply the operator
\begin{equation}
\mathcal{P}(M^{2},m^{2})=\left( 1-M^{2}\frac{d}{dM^{2}}\right)
M^{2}e^{m^{2}/M^{2}},
\end{equation}%
to both sides of the sum rule equality, and subtract remaining conventional
terms in a standard manner.

But the vertex $X_{0}^{\prime }D^{-}K^{+}$ and strong coupling $g^{\prime }$
are also interesting for us. Therefore, we keep the following strategy: we
determine the strong coupling $g$ utilizing the "ground-state+continuum "
scheme and first term in $\Pi ^{\mathrm{Phys}}(p^{2})$. At this stage we
apply the operator $\mathcal{P}(M^{2},\widetilde{m}^{2})$ that singles out
the ground-state term. Afterwards, we use $g$ as an input parameter in
"ground-state+excited-state+continuum " scheme, and by employing full
expression of $\ \Pi ^{\mathrm{Phys}}(p^{2})$ determine the strong coupling $%
g^{\prime }$.

Then the sum rule for $g$ reads
\begin{equation}
g=\frac{m_{c}}{fmf_{D}m_{D}^{2}\widetilde{m}^{2}}\mathcal{P}(M^{2},m^{2})\Pi
^{\mathrm{OPE}}(M^{2},s_{0}),  \label{eq:SC1}
\end{equation}%
whereas for $g^{\prime }$, we obtain
\begin{eqnarray}
g^{\prime } &=&\frac{e^{\widetilde{m}^{\prime 2}/M^{2}}}{f^{\prime
}m^{\prime }\widetilde{m}^{\prime 2}}\left[ \frac{M^{2}m_{c}}{f_{D}m_{D}^{2}}%
\Pi ^{\mathrm{OPE}}(M^{2},s_{0}^{\ast })\right.  \notag \\
&&\left. -gfm\widetilde{m}^{2}e^{-\widetilde{m}^{2}/M^{2}}\right] .
\label{eq:SC2}
\end{eqnarray}

The $K$ meson is characterized by some local matrix elements of different
quark-gluon contents and twists. Having performed numerical computation, we
see that the correlator $\Pi ^{\mathrm{OPE}}(p,q)$ receives contribution
from the two-particle twist-3 element
\begin{equation}
\langle 0|\overline{u}i\gamma _{5}s|K\rangle =\frac{f_{K}m_{K}^{2}}{m_{s}}.
\label{eq:MatElK1}
\end{equation}%
Technical sides of required calculations of the $\Pi ^{\mathrm{OPE}}(p,q)$
in the soft limit were described in Refs.\ \cite{Agaev:2016dev}, hence we
omit further details and write down final formula for the Borel transformed
and subtracted invariant amplitude, which is computed with dimension-$9$
accuracy and given by the formula
\begin{eqnarray}
\Pi ^{\mathrm{OPE}}(M^{2},s_{0}) &=&-\frac{\mu _{K}}{4\pi ^{2}}\int_{%
\mathcal{M}^{2}}^{s_{0}}\frac{ds(m_{c}^{2}-s)^{2}}{s}e^{-s/M^{2}}  \notag \\
&&+\mu _{K}m_{c}\Pi _{\mathrm{NP}}(M^{2}).  \label{eq:DecayCF}
\end{eqnarray}%
The nonperturbative component of the correlation function $\Pi _{\mathrm{NP}%
}(M^{2})$ is determined by the expression%
\begin{eqnarray}
&&\Pi _{\mathrm{NP}}(M^{2})=\frac{2\langle \overline{d}d\rangle }{3}%
e^{-m_{c}^{2}/M^{2}}+\langle \frac{\alpha _{s}G^{2}}{\pi }\rangle \frac{%
m_{c}^{3}}{36M^{4}}  \notag \\
&&\times \int_{0}^{1}\frac{dxe^{-m_{c}^{2}/[M^{2}x(1-x)]}}{x^{3}(1-x)^{3}}-%
\frac{\langle \overline{d}g\sigma Gd\rangle m_{c}^{2}}{3M^{4}}%
e^{-m_{c}^{2}/M^{2}}  \notag \\
&&+\langle \frac{\alpha _{s}G^{2}}{\pi }\rangle \langle \overline{d}d\rangle
\frac{(m_{c}^{2}+3M^{2})\pi ^{2}}{27M^{6}}e^{-m_{c}^{2}/M^{2}}-\langle \frac{%
\alpha _{s}G^{2}}{\pi }\rangle   \notag \\
&&\times \langle \overline{d}g\sigma Gd\rangle \frac{%
(m_{c}^{4}+6M^{2}m_{c}^{2}+6M^{4})\pi ^{2}}{108M^{10}}e^{-m_{c}^{2}/M^{2}},
\label{eq:DecayNPCF}
\end{eqnarray}%
where $\mu _{K}=f_{K}m_{K}^{2}/m_{s}$.

\begin{table}[tbp]
\begin{tabular}{|c|c|}
\hline\hline
Quantity & Value ( $\mathrm{MeV}$) \\ \hline
$m_{D}$ & $1869.65\pm 0.05$ \\
$m_{D^0}$ & $1864.83\pm 0.05$ \\
$m_{K}$ & $493.677\pm 0.016$ \\
$m_{K^0}$ & $497.611\pm 0.013$ \\
$f_{D}=f_{D^0}$ & $212.6 \pm 0.7$ \\
$f_{K}=f_{K^0}$ & $155.7 \pm 0.3$ \\ \hline\hline
\end{tabular}%
\caption{Masses and decay constants of the $D$ and $K$ mesons required for
numerical computations.}
\label{tab:Param}
\end{table}
It is worth noting that the limit $q\rightarrow 0$ is performed in a hard
component of the amplitude. As a result, it does not contain terms $\sim
m_{K}^{2}$ which nevertheless would be small due to $m_{K}^{2}/m^{2}$, $%
m_{K}^{2}/m^{\prime 2}$, $m_{K}^{2}/m_{D}^{2}\ll 1$. In the soft
approximation the mass and decay constant of $K^{+}$ meson through $\mu _{K}$
form the nonperturbative soft factor in $\Pi ^{\mathrm{OPE}}(M^{2},s_{0})$.

Parameters of the mesons $D^{-}$ and $K^{+}$ necessary to calculate $g$ are
removed to Table \ref{tab:Param}. For masses and decay constants of these
particles, we use their values from Ref.\ \cite{PDG:2020}. In numerical
computation of the strong couplings $g$ and $g^{\prime }$ the Borel and
continuum subtraction parameters are chosen as in corresponding mass
analysis. Numerical computations yield
\begin{equation}
g=(1.06\pm 0.27)~\mathrm{GeV}^{-1},  \label{eq:Coupl1}
\end{equation}%
and
\begin{equation}
g^{\prime }=(0.52\pm 0.13)~\mathrm{GeV}^{-1}.  \label{eq:Coupl2}
\end{equation}%
Partial widths of the processes $X_{0}^{(\prime )}\rightarrow D^{-}K^{+\text{
}}$ can be found by means of the expression
\begin{equation}
\Gamma \left[ X_{0}^{(\prime )}\rightarrow D^{-}K^{+\text{ }}\right] =\frac{%
g^{(\prime )2}m_{D}^{2}\lambda ^{(\prime )}}{8\pi }\left( 1+\frac{\lambda
^{(\prime )2}}{m_{D}^{2}}\right) ,  \label{eq:DW}
\end{equation}%
where $\lambda ^{(\prime )}=\lambda \left( m^{(\prime
^{)}},m_{D},m_{K}\right) $ and
\begin{eqnarray}
\lambda \left( a,b,c\right) &=&\frac{1}{2a}\left[ a^{4}+b^{4}+c^{4}\right.
\notag \\
&&\left. -2\left( a^{2}b^{2}+a^{2}c^{2}+b^{2}c^{2}\right) \right] ^{1/2}.
\end{eqnarray}%
Now it is easy to get
\begin{eqnarray}
\Gamma \left[ X_{0}\rightarrow D^{-}K^{+}\right] &=&(64.7\pm 23.3)~\mathrm{%
MeV},  \notag \\
\Gamma \left[ X_{0}^{\prime }\rightarrow D^{-}K^{+}\right] &=&(53.3\pm 18.8)~%
\mathrm{MeV}.  \label{eq:DW12}
\end{eqnarray}

Another processes which form the full widths of the tetraquarks $X_{0}$ and $%
X_{0}^{\prime }$ are decays $X_{0}\rightarrow \overline{D}^{0}K^{0}$ and $%
X_{0}^{\prime }\rightarrow \overline{D}^{0}K^{0}$. Investigation of these
channels runs in accordance with the scheme described above, therefore we
write down only final results: for the strong couplings $G$ and $G^{\prime }$
corresponding to vertices $X_{0}\overline{D}^{0}K^{0}$ and $X_{0}^{\prime }%
\overline{D}^{0}K^{0}$, we find
\begin{equation}
G=(1.14\pm 0.18)~\mathrm{GeV}^{-1},  \label{eq:Coupl3}
\end{equation}%
and
\begin{equation}
G^{\prime }=(0.54\pm 0.11)~\mathrm{GeV}^{-1}.  \label{eq:Coupl4}
\end{equation}%
For partial widths of these decays, we get%
\begin{eqnarray}
\Gamma \left[ X_{0}\rightarrow \overline{D}^{0}K^{0}\right] &=&(74.8\pm
16.7)~\mathrm{MeV},  \notag \\
\Gamma \left[ X_{0}^{\prime }\rightarrow \overline{D}^{0}K^{0}\right]
&=&(56.3\pm 16.2)~\mathrm{MeV}.  \label{eq:DW34}
\end{eqnarray}%
Then full widths of the particles $X_{0}$ and $X_{0}^{\prime }$ are equal to
\begin{eqnarray}
\Gamma _{0} &=&(140\pm 29)~\mathrm{MeV},  \notag \\
\Gamma _{0}^{\prime } &=&(110\pm 25)~\mathrm{MeV},  \label{eq:FullW}
\end{eqnarray}%
respectively.

As is seen, parameters of the tetraquarks $X_{0}$ and $X_{0}^{\prime }$
differ considerably from the mass and width of the resonance $X_{0}(2900)$
measured by the LHCb collaboration.


\section{Discussion and conclusions}

\label{sec:Disc}

In the present paper, we have examined the tetraquark $X_{0}$ and its radial
excitation $X_{0}^{\prime }$ by calculating their masses and widths. The
masses of $X_{0}$ and $X_{0}^{\prime }$ have been computed using the
axial-axial and scalar-scalar type interpolating currents $J(s)$ and $J_{%
\mathrm{S}}(x)$. The widths of these particles have been estimated for the
axial-axial structure.

The diquark-antidiquark state $X_{0}$ is built of the \ four quarks of
different flavors $X_{0}=[ud][\overline{c}\overline{s}]$. Properties of the
ground-state scalar tetraquark with similar content $X_{c}=[su][\overline{c}%
\overline{d}]$ were investigated in Refs. \cite{Agaev:2016lkl,Chen:2016mqt}.
The mass of $X_{c}$ found in Ref. \cite{Agaev:2016lkl} using axial-axial and
scalar-scalar structures is equal to%
\begin{equation}
m_{X_{c}}=(2590\pm 60)~\mathrm{MeV},\ \ \Gamma _{X_{c}}=(63.4\pm 14.2)~%
\mathrm{MeV},  \label{eq:Xc1}
\end{equation}%
and
\begin{equation}
\widetilde{m}_{X_{c}}=(2634\pm 62)~\mathrm{MeV},\ \widetilde{\ \Gamma }%
_{X_{c}}=(57.7\pm 11.6)~\mathrm{MeV,}  \label{eq:Xc2}
\end{equation}%
respectively. The prediction
\begin{equation}
m_{X_{c}}=(2550\pm 90)~\mathrm{MeV}  \label{eq:Xc3}
\end{equation}%
for the mass of the state $X_{c}$ was made also in Ref.\ \cite{Chen:2016mqt}%
. It is worth to emphasize that all of these results were extracted using
the QCD two-point sum rule method, and predictions for the mass of $X_{c}$
from articles \cite{Agaev:2016lkl} and \cite{Chen:2016mqt} almost coincide
with each other. It is also evident that $m$ and $m_{\mathrm{S}}$ in Eq.\ (%
\ref{eq:Result1}) and in Table \ref{tab:Results1} are comparable with
predictions for $m_{X_{c}}$ and $\widetilde{m}_{X_{c}}$ within uncertainties
of computations. Stated differently, the masses of the ground-state
tetraquarks with different internal organizations, but composed of $c$, $s$,
$u$, $d$ quarks vary approximately in limits $2550-2660~\mathrm{MeV}$.

Parameters of the tetraquark $[cs][\overline{u}\overline{d}]$ in the context
of the sum rule approach were recently calculated also in Ref.\ \cite%
{Wang:2020xyc}. Here, for the mass of this particle with the scalar-scalar
(SS) or axial-axial (AA) structures, the author found
\begin{equation}
M_{SS}=(3050\pm 100)~\mathrm{MeV},\ M_{AA}=(2910\pm 120)~\mathrm{MeV.}
\end{equation}%
Because $M_{AA}$ is compatible with the LHCb data, the resonance $%
X_{0}(2900) $ was interpreted there as a ground-state tetraquark $[cs][%
\overline{u}\overline{d}]$. Results of this work differ considerably from
our findings, as well as from prediction made in Ref.\ \cite{Chen:2016mqt}.
The $X_{0}(2900)$ was considered as a radially excited state $\widetilde{X}%
_{c}(2S)$ with $\widetilde{X}_{c}$ being the tetraquark $[ud][\overline{c}%
\overline{s}]$ \cite{He:2020jna}. The mass of the state $\widetilde{X}%
_{c}(2S)$ was estimated there around of $2860\ \mathrm{MeV}$, which is lower
than our results for $X_{0}^{\prime }$ and $X_{\mathrm{S}}^{\prime }$.

Performed analysis allows us to consider tetraquarks $X_{0}^{(\prime )}$ and
$X_{\mathrm{S}}^{(\prime )}$ built of the axial-vector and scalar diquarks
(antidiquarks), respectively, as states that differ from the resonance $%
X_{0}(2900)$ observed by the LHCb collaboration. Therefore, parameters
calculated in the present work are all the more important to search for the
tetraquarks $X_{0}^{(\prime )}$ and $X_{\mathrm{S}}^{(\prime )}$ in various
processes. The masses of states $X_{0}^{(\prime )}$ and $X_{\mathrm{S}%
}^{(\prime )}$ have been extracted with high enough accuracy. Though $%
m^{(\prime )}$ and $m_{\mathrm{S}}^{(\prime )}$ contain uncertainties
typical for all sum rule computations, they provide valuable information on
these exotic mesons. We evaluated also full widths of the tetraquarks $X_{0}$
and $X_{0}^{\prime }$ by considering their decays to pairs of conventional
mesons $D^{-}K^{+}$ and $\overline{D}^{0}K^{0}$. For the particles $X_{0}$
and $X_{0}^{\prime }$ these two processes are their only $S$-wave decay
channels. Other possible modes of the tetraquarks $X_{0}^{(\prime )}$, for
instance, $S$-wave decays $X_{0}^{(\prime )}\rightarrow \overline{D}%
_{0}^{\ast }(2400)^{0}K^{\ast }(1430)$ are kinematically forbidden
processes. Hence, estimates for full widths $\Gamma _{0}^{(\prime )}$ of the
four-quark mesons $X_{0}$ and $X_{0}^{\prime }$ are rather credible.

Our results imply that $X_{0}(2900)$ can not be identified with ground-state
or radially excited scalar tetraquark $[ud][\overline{c}\overline{s}]$. It
seems interpretation of the resonance $X_{0}(2900)$ as hadronic molecules $%
\overline{D}^{\ast 0}K^{\ast 0}$ and $D^{\ast -}K^{\ast +}$, or their some
admixture was correct and overcame successfully this examination.

\begin{widetext}

\appendix*

\section{ The quark propagators and invariant amplitude $\Pi (M^{2},s_{0})$}

\renewcommand{\theequation}{\Alph{section}.\arabic{equation}} \label{sec:App}

In the current article, for the light quark propagator $S_{q}^{ab}(x)$, we
employ the following expression
\begin{eqnarray}
&&S_{q}^{ab}(x)=i\delta _{ab}\frac{\slashed x}{2\pi ^{2}x^{4}}-\delta _{ab}%
\frac{m_{q}}{4\pi ^{2}x^{2}}-\delta _{ab}\frac{\langle \overline{q}q\rangle
}{12}+i\delta _{ab}\frac{\slashed xm_{q}\langle \overline{q}q\rangle }{48}%
-\delta _{ab}\frac{x^{2}}{192}\langle \overline{q}g_{s}\sigma Gq\rangle
\notag \\
&&+i\delta _{ab}\frac{x^{2}\slashed xm_{q}}{1152}\langle \overline{q}%
g_{s}\sigma Gq\rangle -i\frac{g_{s}G_{ab}^{\alpha \beta }}{32\pi ^{2}x^{2}}%
\left[ \slashed x{\sigma _{\alpha \beta }+\sigma _{\alpha \beta }}\slashed x%
\right] -i\delta _{ab}\frac{x^{2}\slashed xg_{s}^{2}\langle \overline{q}%
q\rangle ^{2}}{7776}  \notag \\
&&-\delta _{ab}\frac{x^{4}\langle \overline{q}q\rangle \langle
g_{s}^{2}G^{2}\rangle }{27648}+\cdots .  \label{eq:qprop}
\end{eqnarray}%
For the heavy quark $Q=c$, we use the propagator $S_{Q}^{ab}(x)$
\begin{eqnarray}
&&S_{Q}^{ab}(x)=i\int \frac{d^{4}k}{(2\pi )^{4}}e^{-ikx}\Bigg \{\frac{\delta
_{ab}\left( {\slashed k}+m_{Q}\right) }{k^{2}-m_{Q}^{2}}-\frac{%
g_{s}G_{ab}^{\alpha \beta }}{4}\frac{\sigma _{\alpha \beta }\left( {\slashed %
k}+m_{Q}\right) +\left( {\slashed k}+m_{Q}\right) \sigma _{\alpha \beta }}{%
(k^{2}-m_{Q}^{2})^{2}}  \notag \\
&&+\frac{g_{s}^{2}G^{2}}{12}\delta _{ab}m_{Q}\frac{k^{2}+m_{Q}{\slashed k}}{%
(k^{2}-m_{Q}^{2})^{4}}+\frac{g_{s}^{3}G^{3}}{48}\delta _{ab}\frac{\left( {%
\slashed k}+m_{Q}\right) }{(k^{2}-m_{Q}^{2})^{6}}\left[ {\slashed k}\left(
k^{2}-3m_{Q}^{2}\right) +2m_{Q}\left( 2k^{2}-m_{Q}^{2}\right) \right] \left(
{\slashed k}+m_{Q}\right) +\cdots \Bigg \}.  \notag \\
&&  \label{eq:Qprop}
\end{eqnarray}

Here, we have used the short-hand notations
\begin{equation}
G_{ab}^{\alpha \beta }\equiv G_{A}^{\alpha \beta }\lambda _{ab}^{A}/2,\ \
G^{2}=G_{\alpha \beta }^{A}G_{A}^{\alpha \beta },\ G^{3}=f^{ABC}G_{\alpha
\beta }^{A}G^{B\beta \delta }G_{\delta }^{C\alpha },
\end{equation}%
where $G_{A}^{\alpha \beta }$ is the gluon field strength tensor, $\lambda
^{A}$ and $f^{ABC}$ are the Gell-Mann matrices and structure constants of
the color group $SU_{c}(3)$, respectively. The indices $A,B,C$ run in the
range $1,2,\ldots 8$.

The invariant amplitude $\Pi (M^{2},s_{0})$, obtained using the
interpolating current $J(x)$ from Eq.\ (\ref{eq:CR1}), after the Borel
transformation and subtraction procedures is given by the expression%
\begin{equation*}
\Pi (M^{2},s_{0})=\int_{\mathcal{M}^{2}}^{s_{0}}ds\rho ^{\mathrm{OPE}%
}(s)e^{-s/M^{2}}+\Pi (M^{2}),
\end{equation*}%
where the spectral density $\rho ^{\mathrm{OPE}}(s)$ and the function $\Pi
(M^{2})$ are determined by formulas
\begin{equation}
\rho ^{\mathrm{OPE}}(s)=\rho ^{\mathrm{pert.}}(s)+\sum_{N=3}^{8}\rho ^{%
\mathrm{DimN}}(s),\ \ \Pi (M^{2})=\sum_{N=6}^{15}\Pi ^{\mathrm{DimN}}(M^{2}),
\label{eq:A1}
\end{equation}%
respectively. The components of $\rho ^{\mathrm{OPE}}(s)$ and $\Pi (M^{2})$
are given by the expressions%
\begin{equation}
\rho ^{\mathrm{DimN}}(s)=\int_{0}^{1}d\alpha \rho ^{\mathrm{DimN}}(s,\alpha
),\ \ \Pi ^{\mathrm{DimN}}(M^{2})=\int_{0}^{1}d\alpha \Pi ^{\mathrm{DimN}%
}(M^{2},\alpha ).  \label{eq:A2}
\end{equation}%
In Eq.\ (\ref{eq:A2}) the variable $\alpha $ is the Feynman parameter.

The perturbative and nonperturbative components of the spectral density $%
\rho ^{\mathrm{pert.}}(s,\alpha )$ and $\rho ^{\mathrm{Dim3(4,5,6,7,8)}%
}(s,\alpha )$ \ are given by the following expressions
\begin{equation}
\rho ^{\mathrm{pert.}}(s,\alpha )=\frac{\left[ m_{c}^{2}-s(1-\alpha )\right]
^{3}\alpha ^{3}\Theta (L)}{1536\pi ^{6}(\alpha -1)^{3}}\left[
4m_{c}m_{s}+m_{c}^{2}\alpha +3s\alpha (1-\alpha )\right] ,
\end{equation}%
\begin{equation}
\rho ^{\mathrm{Dim3}}(s,\alpha )=-\frac{\langle \overline{s}s\rangle \Theta
(L)}{36\pi ^{2}(\alpha -1)^{2}}\alpha ^{2}\left[ m_{c}^{2}-s(1-\alpha )%
\right] \left[ m_{c}^{3}+2m_{c}^{2}m_{s}(\alpha -1)+m_{c}s(\alpha
-1)+4m_{s}s(\alpha -1)^{2}\right]
\end{equation}%
\begin{eqnarray}
&&\rho ^{\mathrm{Dim4}}(s,\alpha )=-\frac{\langle \alpha _{s}G^{2}/\pi
\rangle \Theta (L)}{9\cdot 2^{9}\pi ^{4}(\alpha -1)^{3}}\alpha ^{2}\left[
6s^{2}(\alpha -1)^{3}(5\alpha -6)+m_{c}^{3}m_{s}(-9+9\alpha -8\alpha
^{2})+m_{c}^{4}\left( 18-33\alpha \right. \right.   \notag \\
&&\left. \left. +19\alpha ^{2}\right) +m_{c}m_{s}s\left( 9-22\alpha
+25\alpha ^{2}-12\alpha ^{3}\right) +3m_{c}^{2}s\left( -18+51\alpha
-50\alpha ^{2}+17\alpha ^{3}\right) \right] ,
\end{eqnarray}%
\begin{equation}
\rho ^{\mathrm{Dim5}}(s,\alpha )=\frac{\langle \overline{s}g_{s}\sigma
Gs\rangle \Theta (L)}{96\pi ^{4}(\alpha -1)}\alpha \left[
3m_{c}^{3}+4m_{c}^{2}m_{s}(\alpha -1)+3m_{c}s(\alpha -1)+6sm_{s}(\alpha
-1)^{2}\right] ,
\end{equation}%
\begin{eqnarray}
&&\rho ^{\mathrm{Dim6}}(M^{2},\alpha )=-\frac{\Theta (L)}{405\cdot 2^{9}\pi
^{6}(\alpha -1)^{3}}\left\{ 27\langle g_{s}^{3}G^{3}\rangle m_{c}^{2}\alpha
^{5}+34560\langle \overline{d}d\rangle \langle \overline{u}u\rangle \pi
^{4}(\alpha -1)^{3}\left[ -2m_{c}m_{s}+2m_{c}^{2}\alpha \right. \right.
\notag \\
&&\left. +3s\alpha (\alpha -1)\right] +320g_{s}^{2}\langle \overline{d}%
d\rangle ^{2}\pi ^{2}(\alpha -1)^{3}\left[ -m_{c}m_{s}+4m_{c}^{2}\alpha
+6s\alpha (\alpha -1)\right] +320g_{s}^{2}\pi ^{2}(\alpha -1)^{3}  \notag \\
&&\left. \times \left[ -m_{c}m_{s}\langle \overline{u}u\rangle ^{2}+\left(
\langle \overline{s}s\rangle ^{2}+\langle \overline{u}u\rangle ^{2}\right)
\left( 4m_{c}^{2}\alpha +6s\alpha (\alpha -1)\right) \right] \right\} ,
\end{eqnarray}%
\begin{equation}
\rho ^{\mathrm{Dim7}}(M^{2},\alpha )=\frac{\langle \alpha _{s}G^{2}/\pi
\rangle \langle \overline{s}s\rangle \Theta (L)}{288\pi ^{2}(\alpha -1)^{2}}%
\left[ 3m_{s}\alpha (\alpha -1)^{2}+m_{c}\left( 2-7\alpha +5\alpha
^{2}-2\alpha ^{3}\right) \right] ,
\end{equation}%
\begin{equation}
\rho ^{\mathrm{Dim8}}(M^{2},\alpha )=\frac{\langle \alpha _{s}G^{2}/\pi
\rangle ^{2}\alpha +96\langle \overline{d}g_{s}\sigma Gd\rangle \langle
\overline{u}u\rangle (\alpha -1)}{1152\pi ^{2}}\Theta (L),
\end{equation}

Components of the function $\Pi (M^{2})$ are:%
\begin{eqnarray}
\Pi ^{\mathrm{Dim6}}(M^{2},\alpha ) &=&-\frac{\langle g_{s}^{3}G^{3}\rangle
m_{c}^{3}\alpha ^{3}}{45\cdot 2^{10}M^{2}\pi ^{6}(\alpha -1)^{5}}\exp \left[
-\frac{m_{c}^{2}}{M^{2}(1-\alpha )}\right] \left[ m_{c}^{3}\alpha (2+\alpha
)+4m_{c}^{2}m_{s}(2\alpha -1)-8m_{s}M^{2}(\alpha ^{2}-1)\right.   \notag \\
&&\left. -m_{c}M^{2}\alpha (\alpha ^{2}+\alpha -2)\right] ,
\end{eqnarray}%
\begin{equation}
\Pi ^{\mathrm{Dim7}}(M^{2},\alpha )=\frac{\langle \alpha _{s}G^{2}/\pi
\rangle \langle \overline{s}s\rangle m_{c}^{2}\alpha ^{2}}{288M^{2}\pi
^{2}(\alpha -1)^{3}}\exp \left[ -\frac{m_{c}^{2}}{M^{2}(1-\alpha )}\right] %
\left[ m_{c}^{2}m_{s}+(m_{c}-m_{s})M^{2}(\alpha -1)\right] ,
\end{equation}%
\begin{eqnarray}
\Pi _{1}^{\mathrm{Dim8}}(M^{2},\alpha ) &=&\frac{\langle \alpha
_{s}G^{2}/\pi \rangle ^{2}m_{c}\alpha }{9\cdot 2^{9}M^{2}\pi ^{2}(\alpha
-1)^{3}}\exp \left[ -\frac{m_{c}^{2}}{M^{2}(1-\alpha )}\right] \left[
2m_{c}^{2}m_{s}(\alpha -1)+m_{c}^{3}\alpha -m_{c}M^{2}\alpha (\alpha
-1)\right.   \notag \\
&&\left. -2m_{s}M^{2}(\alpha ^{2}-1)\right] , \\
\Pi _{2}^{\mathrm{Dim8}}(M^{2}) &=&\frac{\langle \overline{d}g_{s}\sigma
Gd\rangle \langle \overline{u}u\rangle m_{c}m_{s}}{12\pi ^{2}}\exp \left[ -%
\frac{m_{c}^{2}}{M^{2}}\right] ,
\end{eqnarray}%
\begin{eqnarray}
&&\Pi _{1}^{\mathrm{Dim9}}(M^{2},\alpha )=-\frac{1}{135\cdot 2^{7}M^{6}\pi
^{4}(\alpha -1)^{5}}\exp \left[ -\frac{m_{c}^{2}}{M^{2}(1-\alpha )}\right]
\left\{ 3\langle g_{s}^{3}G^{3}\rangle \langle \overline{s}s\rangle
m_{c}^{3}\alpha ^{2}\left[ m_{c}^{2}M^{2}(2-4\alpha )+8M^{4}(\alpha
-1)\right. \right.   \notag \\
&&\left. +m_{c}^{3}m_{s}(2+\alpha )\right] +5\langle \alpha _{s}G^{2}/\pi
\rangle \langle \overline{s}g_{s}\sigma Gs\rangle M^{2}\pi ^{2}(\alpha
-1)^{2}\left[ -3m_{c}^{2}m_{s}M^{2}(\alpha -1)^{2}+3m_{s}M^{4}(\alpha
-1)^{3}+4m_{c}^{4}m_{s}\alpha \right.   \notag \\
&&\left. \left. +6m_{c}^{3}M^{2}\alpha (\alpha -1)+3m_{c}M^{4}(3-4\alpha
+3\alpha ^{2}-2\alpha ^{3})\right] \right\} , \\
&&\Pi _{2}^{\mathrm{Dim9}}(M^{2})=\frac{\langle \overline{s}s\rangle }{%
486M^{2}\pi ^{2}}\exp \left[ -\frac{m_{c}^{2}}{M^{2}}\right] \left\{
g_{s}^{2}\left[ \langle \overline{d}d\rangle ^{2}+\langle \overline{u}%
u\rangle ^{2}\right] \left[ m_{c}^{2}m_{s}+(m_{c}-m_{s})M^{2}\right]
+54\langle \overline{u}u\rangle \langle \overline{d}d\rangle \pi ^{2}\right.
\notag \\
&&\left. \times \left[ 4m_{c}M^{2}+m_{s}(m_{c}^{2}-M^{2})\right] \right\} ,
\end{eqnarray}%
\begin{eqnarray}
&&\Pi _{1}^{\mathrm{Dim10}}(M^{2},\alpha )=\frac{g_{s}^{2}\langle \alpha
_{s}G^{2}/\pi \rangle }{729\cdot 2^{6}M^{4}\pi ^{2}(\alpha -1)^{3}}\exp %
\left[ -\frac{m_{c}^{2}}{M^{2}(1-\alpha )}\right] \left\{ -4\langle
\overline{u}u\rangle ^{2}m_{c}^{3}m_{s}(\alpha -1)+8\langle \overline{u}%
u\rangle ^{2}m_{c}m_{s}M^{2}(\alpha -1)\right.   \notag \\
&&+\left[ 2\langle \overline{s}s\rangle ^{2}+\langle \overline{u}u\rangle
^{2}\right] \left[ -3m_{c}^{2}M^{2}(\alpha -1)^{2}+3M^{4}(\alpha -1)^{3}%
\right] +8m_{c}^{4}\alpha \left[ \langle \overline{s}s\rangle ^{2}+\langle
\overline{u}u\rangle ^{2}\right] +\langle \overline{d}d\rangle ^{2}\left[
-4m_{c}^{3}m_{s}(\alpha -1)\right.   \notag \\
&&\left. \left. +8m_{c}m_{s}M^{2}(\alpha -1)-3m_{c}^{2}M^{2}(\alpha
-1)^{2}+3M^{4}(\alpha -1)^{3}+8m_{c}^{4}\alpha \right] \right\} , \\
&&\Pi _{2}^{\mathrm{Dim10}}(M^{2})=\frac{1}{243\cdot 2^{6}M^{4}\pi ^{2}}\exp %
\left[ -\frac{m_{c}^{2}}{M^{2}}\right] \left\{ 162\langle \overline{u}%
g_{s}\sigma Gu\rangle \langle \overline{d}g_{s}\sigma Gd\rangle \left(
2m_{c}^{3}m_{s}+m_{c}^{2}M^{2}-M^{4}\right) \right.   \notag \\
&&\left. +\langle \alpha _{s}G^{2}/\pi \rangle \left[ g_{s}^{2}(\langle
\overline{d}d\rangle ^{2}+\langle \overline{u}u\rangle
^{2})M^{2}(m_{c}^{2}+m_{c}m_{s}-M^{2})+144\pi ^{2}\langle \overline{u}%
u\rangle \langle \overline{d}d\rangle \left(
2m_{c}^{3}m_{s}+m_{c}^{2}M^{2}-M^{4}\right) \right] \right\} ,
\end{eqnarray}%
and%
\begin{eqnarray}
&&\Pi _{1}^{\mathrm{Dim11}}(M^{2},\alpha )=-\frac{m_{c}}{405\cdot
2^{8}M^{8}\pi ^{4}(\alpha -1)^{5}}\exp \left[ -\frac{m_{c}^{2}}{%
M^{2}(1-\alpha )}\right] \left\{ -20\langle \alpha _{s}G^{2}/\pi \rangle
^{2}\langle \overline{s}s\rangle (m_{c}^{2}-2M^{2})M^{4}\pi ^{4}(\alpha
-1)^{3}\right.   \notag \\
&&+3\langle g_{s}^{3}G^{3}\rangle \langle \overline{s}g_{s}\sigma Gs\rangle
m_{c}\alpha \left[ 2m_{c}^{4}m_{s}(2+\alpha )+6m_{c}^{3}M^{2}(2\alpha
-1)-3m_{s}M^{4}(\alpha -1)^{2}(5\alpha -7)+12m_{c}M^{4}(3-4\alpha +\alpha
^{2})\right.   \notag \\
&&\left. \left. -3m_{c}^{2}m_{s}M^{2}(1-4\alpha +3\alpha ^{2})\right]
\right\} , \\
&&\Pi _{2}^{\mathrm{Dim11}}(M^{2})=-\frac{m_{c}}{729\cdot 2^{7}M^{6}\pi ^{2}}%
\exp \left[ -\frac{m_{c}^{2}}{M^{2}}\right] \left\{ 16g_{s}^{2}\langle
\overline{s}g_{s}\sigma Gs\rangle \left[ \langle \overline{d}d\rangle
^{2}+\langle \overline{u}u\rangle ^{2}\right]
m_{c}(2m_{c}^{2}m_{s}+3m_{c}M^{2}+6m_{s}M^{2})\right.   \notag \\
&&\left. +27\langle \alpha _{s}G^{2}/\pi \rangle ^{2}\langle \overline{s}%
s\rangle M^{4}\pi ^{2}+3456\langle \overline{s}g_{s}\sigma Gs\rangle \langle
\overline{d}d\rangle \langle \overline{u}u\rangle \pi
^{2}m_{c}(2m_{c}^{2}m_{s}+9m_{c}M^{2}+6m_{s}M^{2})\right\} .
\end{eqnarray}

In expressions above, $\Theta (z)$ is Unit Step function. We have used also
the following short-hand notations%
\begin{equation}
L\equiv L(s,\alpha )=s\alpha (1-\alpha )-m_{c}^{2}\alpha .
\end{equation}

\end{widetext}

\end{document}